\pdfoutput=1

\documentclass[a4paper]{article}

\usepackage{atmohead2013}
\usepackage[english]{babel}

\def\hc#1{\leavevmode\hbox to \hsize{\hss #1\hss}\leavevmode}
\def\includefigure#1{\hc{\resizebox{\columnwidth}{!}{\includegraphics{#1}}}}

\title{Simulations of detector arrays and the impact of atmospheric parameters}

\shorttitle{Simulations of detector arrays and the impact of atmospheric parameters}

\authors{
Konrad Bernl\"ohr$^{1,2}$
}

\afiliations{
$^1$ Max-Planck-Institut f\"ur Kernphysik, Postfach 103980, 69029 Heidelberg \\
$^2$ Humboldt-Universit\"at zu Berlin, Institut f\"ur Physik, Newtonstra{\ss}e 15, 12489 Berlin \\
}

\email{Konrad.Bernloehr@mpi-hd.mpg.de}

\abstract{%
In Monte-Carlo simulations of gamma-ray or cosmic-ray detector arrays on the ground 
(here mainly arrays of imaging atmospheric Cherenkov telescopes), the atmosphere enters in 
several ways: in the development of the particle showers, in the emission of 
light by shower particles, and in the propagation of Cherenkov light 
(or fluorescence light or of particles) down to ground level. 
Relevant parameters and their typical impact on energy scale and so on
are discussed here.
}


\begin{document}
\maketitle

\section{Introduction}

Ground-based gamma-ray and cosmic-ray detectors make use of the atmosphere
as a part of their instruments. Incoming energetic particles (neglecting neutrinos
or exotic particles for the purpose of this paper) will interact with the atmosphere,
produce secondary particles and initiate a particle cascade or extensive air shower.
The level of inclusion of the atmosphere into the
instrument differs between different detection techniques. Particle detector arrays
are affected by the development of extensive air showers. Beyond the atmospheric
overburden on ground, the atmospheric profile will also matter, due to the
competition between interactions and decays of secondary unstable particles
produced in interactions. Since most of the development of air showers happens
at altitudes of up to a few ten kilometers of altitude, the change of
atmospheric composition beyond about 80 km altitude has, fortunately, 
no significant impact onto the air shower development and detection.

For atmospheric Cherenkov detectors as well as for air fluorescence detectors,
the atmosphere enters at two additional levels. The first is the emission process
and the second is the propagation of the emitted light down to the photo-detectors.
The Cherenkov and fluorescence processes depend in different ways on changes
in density profiles. For the fluorescence technique, there is a competition between 
fluorescence light emission and collisional de-excitation, after the nitrogen
molecules get excited by the charged particles in the air shower. The higher
efficiency of light emission is balanced basically by lower excitation in lower
density air. As a consequence, the fluorescence process depends relatively
weakly on the air density profile, except for the shower development itself.
The Cherenkov emission process is more strongly impacted by the density
profile. Higher density, resulting in an almost proportionally enhanced
refractivity (index of refraction minus one, $n-1$), results in enhanced emission
as well as lowering the minimum particle velocity necessary for emission.
It also increases the Cherenkov emission angle and spreading out the 
light on the ground over a larger area.

Both the Cherenkov and fluorescence techniques are affected by the
light propagation, by absorption as well as by scattering. The absorption
processes are not just affected by the main composition components but
also affected by trace gases, like ozone or water vapour. At that stage the
detailed composition profile thus enters the game. Scattering includes
Rayleigh scattering - thus directly linked to the density profile - and
Mie scattering on aerosols (also resulting in light absorption). As a consequence,
these detection techniques are affected by the aerosol density,
chemical composition, size distributions -- and all of these depend on
altitude and change with time.

Simulations of air showers and detector response usually need some
simplifications, in order to run in an efficient way. This paper will also try
to point out some often-used simplifications which should be 
applied with care.

Since this paper is focusing on the impact of atmospheric parameters,
a number of other site-related topics will be omitted, including the
impact of the night-sky background (NSB) on Cherenkov or fluorescence detectors
and the impact of the geomagnetic field.

\section{Shower development and density profile}

\begin{figure}[t]
  \centering
  \includefigure{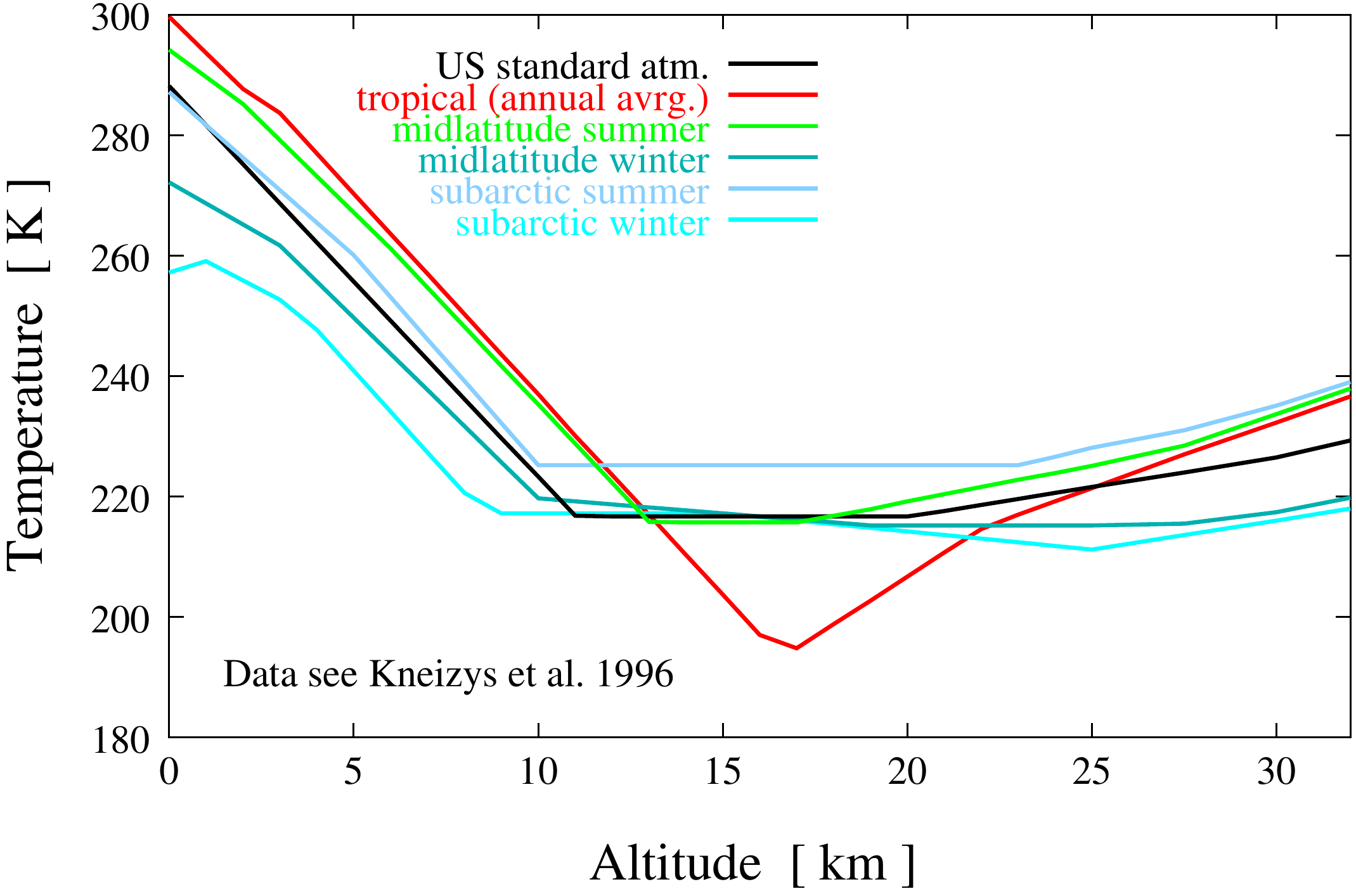}
  \includefigure{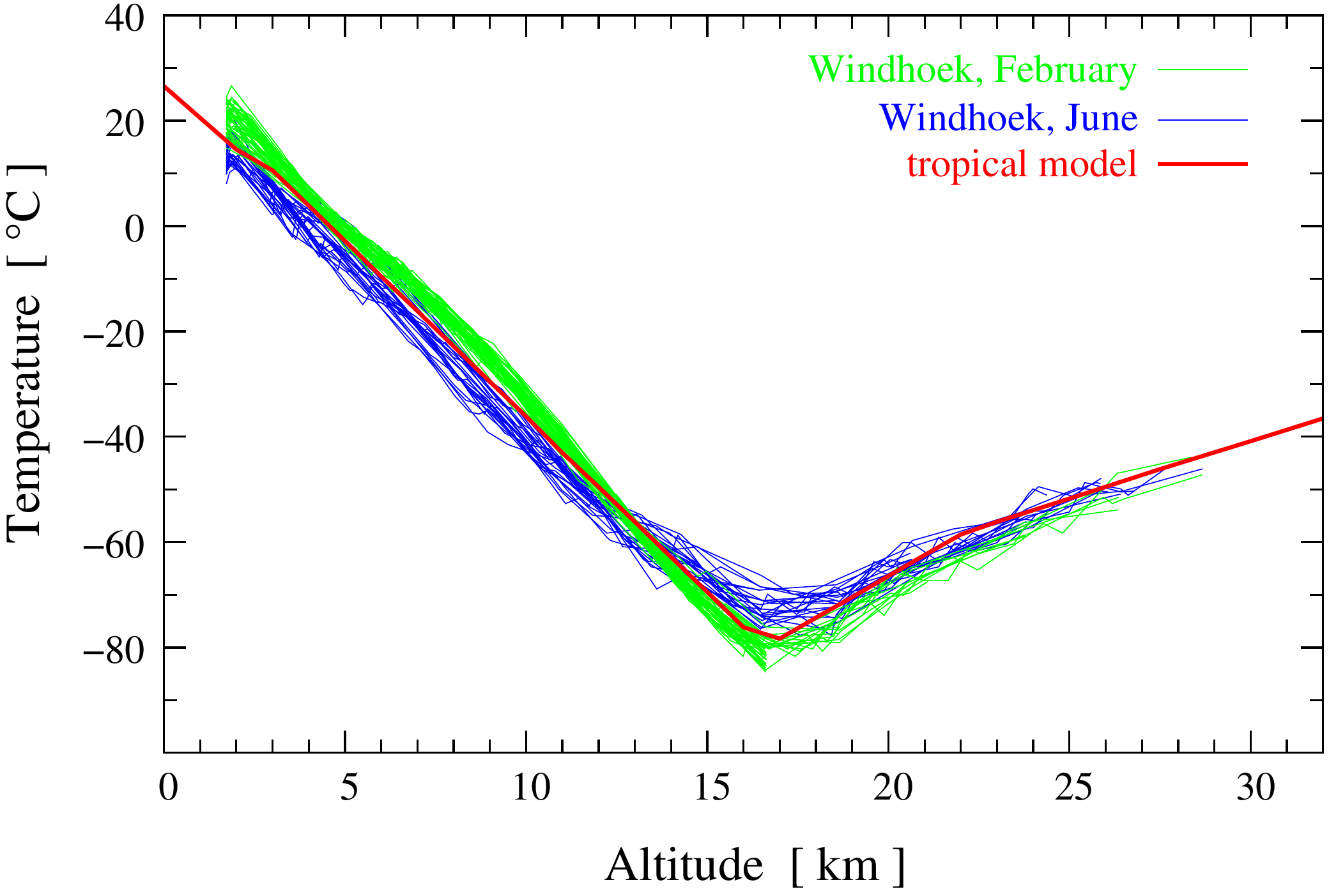}
  \caption{Examples of atmospheric profiles (here: temperature profiles).
    Top: MODTRAN \cite{modtran} model profiles. Bottom: Seasonal extreme
    profiles from radiosonde data taken near Windhoek, Namibia, compared
    with the tropical model.}
  \label{fig:hess-profile}
\end{figure}

Ground-level measurements of the pressure and gravity can tell the total
atmospheric overburden as the most important atmospheric parameter for
particle detector arrays. This will not be enough for a detailed simulation
of the shower development, in particular for the electron/muon ratio.
This is due to the competition between interaction and decay for particles
like kaons and charged pions. Different atmospheric density profiles can
therefore result in differences in the longitudinal shower development, even
when this is expressed as a function of atmospheric depth traversed
(in g/cm$^2$, counted from the top of the atmosphere). 
Even without such differences, different atmospheric 
profiles can result in different lateral distributions of particles on the ground
because the relation between atmospheric depth and altitude is changed.
Simulations will typically be set up to match an average (or seasonal average)
of the density profile at the detector site. These profiles may be pure model
profiles, e.g. from MODTRAN \cite{modtran} or NRLMSISE-00 \cite{nrlmsise} or
from radiosonde data complemented by models or other data beyond the
altitude range covered by the radiosondes. For latitudes up to about 30 degrees
the MODTRAN tropical model is generally a good description and seasonal
variations typically small (see Figure \ref{fig:hess-profile}, resulting changes
in Cherenkov light density below 5 percent) 
while local and seasonal variations at higher latitudes are larger
(resulting in changes of 10 percent or more).

Since simulations are carried out for average profiles and not for instantaneous profiles, even
if these were available,  corrections for the impact of slightly different
profiles may improve the resulting accuracy, .e.g based on ground-level
measurements like pressure and temperature or derived from weather models
like the altitude of some pressure levels. 
Frequent simplifications in simulation programs include piece-wise
exponential or linear density profiles as a function of altitude,
and in most cases also an altitude-independent composition.
The former can sometimes result in artifacts and needs some
care while the latter is not considered a problem since the first interaction
is typically well below 80~km where the composition
starts to change.

\begin{figure}[t]
  \centering
  \includegraphics[width=0.4\textwidth]{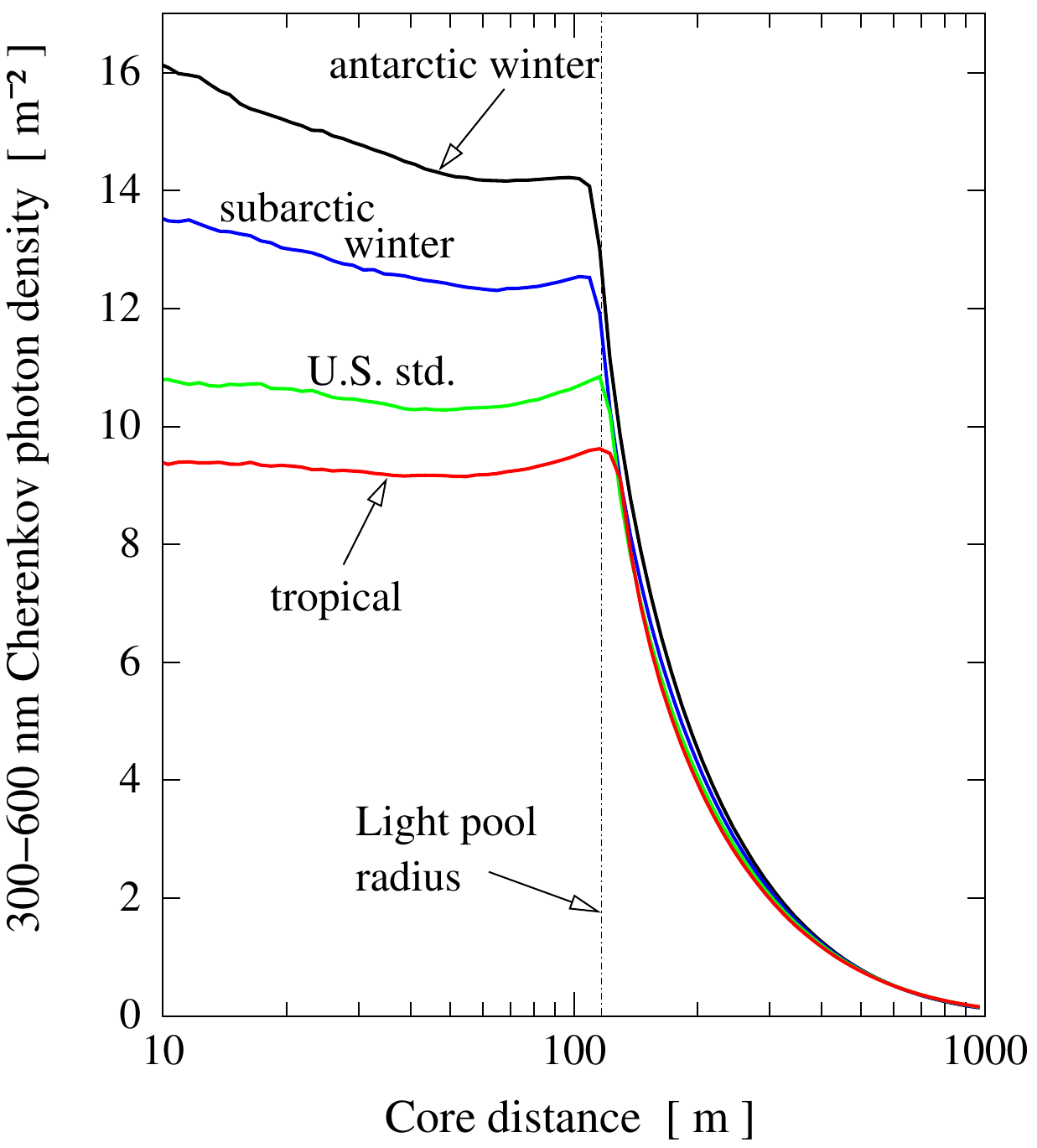}
  \caption{Comparison of average Cherenkov light density on the ground
    for vertical 100 GeV gamma-ray showers at an observation level of 2200~m,
    for different atmospheric profiles from Figure \ref{fig:hess-profile} (top).}
  \label{fig:comp-profiles}
\end{figure}

\section{The Cherenkov emission process}

\begin{figure}[t]
  \centering
  \includegraphics[width=0.43\textwidth]{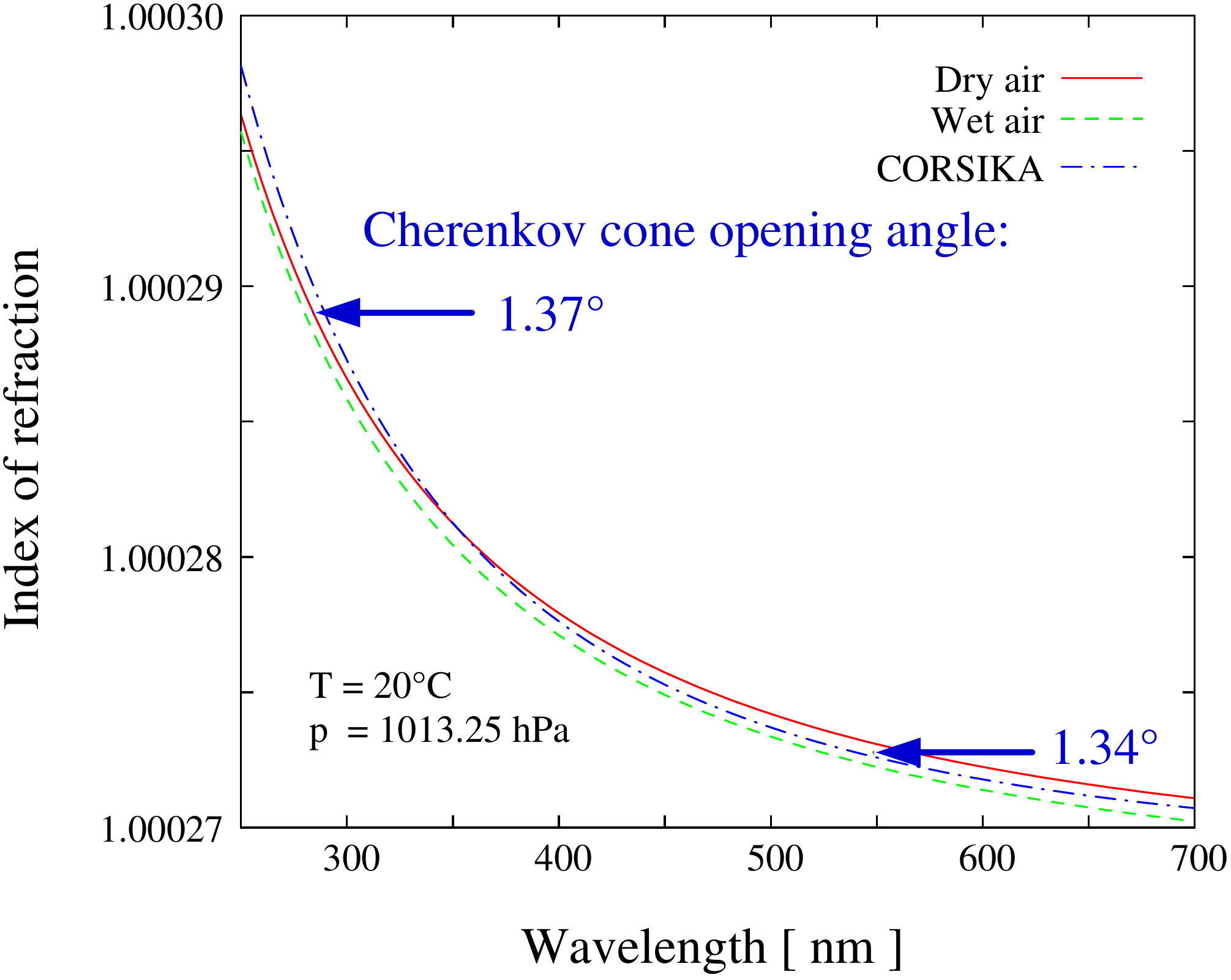}
  \caption{The index of refraction in dry and wet air at normal conditions
     as a function of the wavelength. In addition to the accurate calculations
     \cite{ciddor1996}
     the simplified wavelength dependence \cite{bernlohr2008} available in the CORSIKA 
     air shower simulation package \cite{corsika} is also shown.}
  \label{fig:refidx}
\end{figure}

The atmospheric Cherenkov technique is affected to a large extent by
the atmospheric profiles through the longitudinal shower development
(see Figure \ref{fig:comp-profiles}).
The next obvious impact of atmospheric parameters on the Cherenkov light
emission is by the index of refraction $n$, through the Cherenkov cone opening
angle $\cos \theta=1/(n\beta)$ with $\beta=v/c$ standing for the velocity $v$
of the emitting particle. The index of refraction also enters into the amount of
Cherenkov light emitted per unit path length and unit wavelength being
proportional to $\sin \theta$. The refractivity $n-1$ is to good approximation
proportional to the air density, but for improved accuracy additional corrections are needed
for composition (in particular the water vapour partial pressure) and for
wavelength. The impact of both of these is illustrated in Figure~\ref{fig:refidx}.

The most important simplification in simulations is perhaps to
neglect the wavelength dependence of the index of refraction.
The CORSIKA program \cite{corsika}, for example, uses by default a fixed index
of refraction corresponding to a wavelength of 400 nanometers
but provides the option to include the wavelength dependence
\cite{bernlohr2008}, although at a significant cost in terms of efficiency.
This is related to the reduction of photons tracked down to the
telescopes, in CORSIKA by means of grouping them into 
{\em photon bunches}, to a level just low enough that they would usually not result
in more than one photo-electron in the detector simulation.
With wavelength-independent index of refraction, this `low enough'
is related to the inverse of the average photon detection efficiency over the whole
wavelength range while in the wavelength-dependent case it is related
to the inverse of the peak photon detection efficiency.

\section{Dependence on site altitude}

\begin{figure}[t]
  \centering 
  \includegraphics[width=0.34\textwidth]{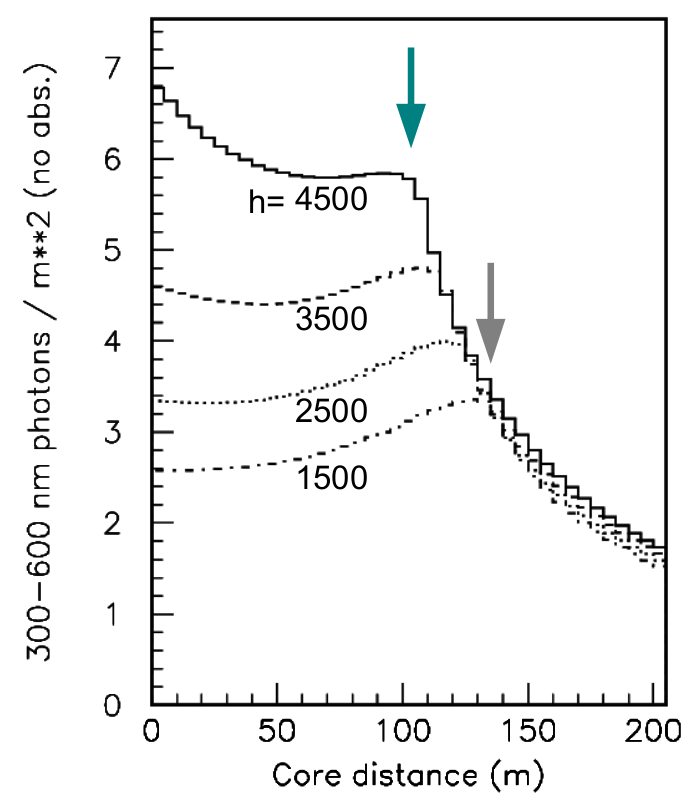}
  \caption{The dependence of the lateral density of Cherenkov photons on the ground
    from vertical showers initiated by 30 GeV gamma rays, as a function of
    core distance for different observation altitudes $h$. No field-of-view restrictions
    and no extinction of Cherenkov light are applied here.
    In general, the larger photon density in the so-called light pool (see arrows)
    at high altitude is because the light is spread out over a smaller area than
    at lower altitudes.
    The resulting photon density at large core distances (not shown here) will
    be smaller at higher altitude. The peak in photon density at small core distances,
    only visible for high altitudes,
    is mainly due to particles penetrating down to observation level.}
  \label{fig:alt-iact}
\end{figure}

Observation of air showers with imaging atmospheric Cherenkov telescopes (IACTs)
depends quite significantly on the observation altitude -- even though
this dependence is much weaker than the altitude dependence for
particle detector arrays. The primary altitude dependence for IACTs is due
to their distance from the shower maximum (see Figure \ref{fig:alt-iact}).
At higher altitudes, the telescopes are closer to the emission region and the
emitted light thus less diluted at the observation level, resulting in a larger
density of photons over a smaller area on the ground. At high altitudes, the
energy threshold of IACTs will thus be lower. 

At the same time, IACTs at
high altitude will see larger fluctuations in the shower images because of
random particles penetrating down to or close to ground level. 
As a result, it becomes increasingly difficult to distinguish between
gamma- (or electron-) initiated electromagnetic showers and the
usually more irregular hadron showers initiated by protons or nuclei
(poor gamma-hadron discrimination).

Also the instrument field-of-view (FoV) may be too small to image all of the
Cherenkov light, in particular at a large impact parameter (distance to
the shower axis). The necessary FoV (and thus telescope cost) to 
completely catch a shower image is increasing with altitude.
For a fixed FoV of telescopes in an IACT array, showers are generally
seen in fewer telescopes at a high-altitude site, resulting in a poorer
angular and energy reconstruction and, again, in poor gamma-hadron discrimination.

As a consequence of these effects, high-altitude sites are generally only
preferable for observations close to the energy threshold of the
instrument. For energies well above threshold, an IACT array at a lower
altitude will result in better angular and energy resolution, in a
larger effective detection area and, in particular, in a better
rejection of the cosmic-ray background.

\section{Extinction and scattering of Cherenkov light or fluorescence light}

A significant part of the Cherenkov light or fluorescence light produced by
air showers is lost on the way to the ground. As long as light scattered into
the detector FoV within the signal readout period is of little relevance, this
loss is just the extinction. The extinction includes absorption by a number
of molecules as well as Rayleigh scattering and Mie scattering on aerosols.
Among the absorbing molecules, ozone and normal oxygen are most prominent.
For ozone, the main absorption bands are the Hartly bands in the 200-300 nm
range and the Huggins bands up to 340~nm; the Chappuis bands near 600 nm
being much weaker.  Oxygen absorption is most relevant in the Herzberg
continuum below 242 nm and in the Herzberg band around 260 nm. In addition,
there is some absorption on water vapour (see \cite{bernlohr2000}).

For typical Cherenkov detector, using photomultiplier tubes with borosilicate
windows, the impact of the highly variable tropospheric ozone profile on the
signal from air showers is only at the level of one or two percent. For detectors
with higher UV response, it can be really significant, in particular when
muon rings are used for calibration purposes. Calibration of Cherenkov
detectors with significant photon detection efficiency below 280 nm
will therefore require much more intense atmospheric monitoring than
with typical current devices. For the detection of fluorescence light, most
of which is emitted in the 300-400 nm range, tropospheric ozone is not
an issue. The stratospheric ozone layer, on the other hand, is not relevant
since at this altitude the shower development has usually not even started.

The molecular Rayleigh scattering can be accurately calculated from the
density profile and included in the simulations. The main uncertainty in
the extinction of Cherenkov or fluorescence light is generally the Mie
scattering on aerosols. Even under fairly good observation conditions,
the aerosol optical depth can easily vary by 0.1, resulting in 10 percent
more or less light in the detector. The vertical profile of these
aerosols is best monitored with Lidars, if possible without interfering
too much with observations then best along the line of sight of the
telescopes. The total extinction is more reliably obtained from
star light, and is an important cross-check to the atmospheric
transmission tables required in the simulation of the light yield
from the air showers. Since aerosols (including hydrosols) can change
with temperature, day-time measurements (e.g. based on scattered
sun light) can be biased -- simulations are better based on
measurements during night time.

\begin{figure}[t]
  \centering 
  \includefigure{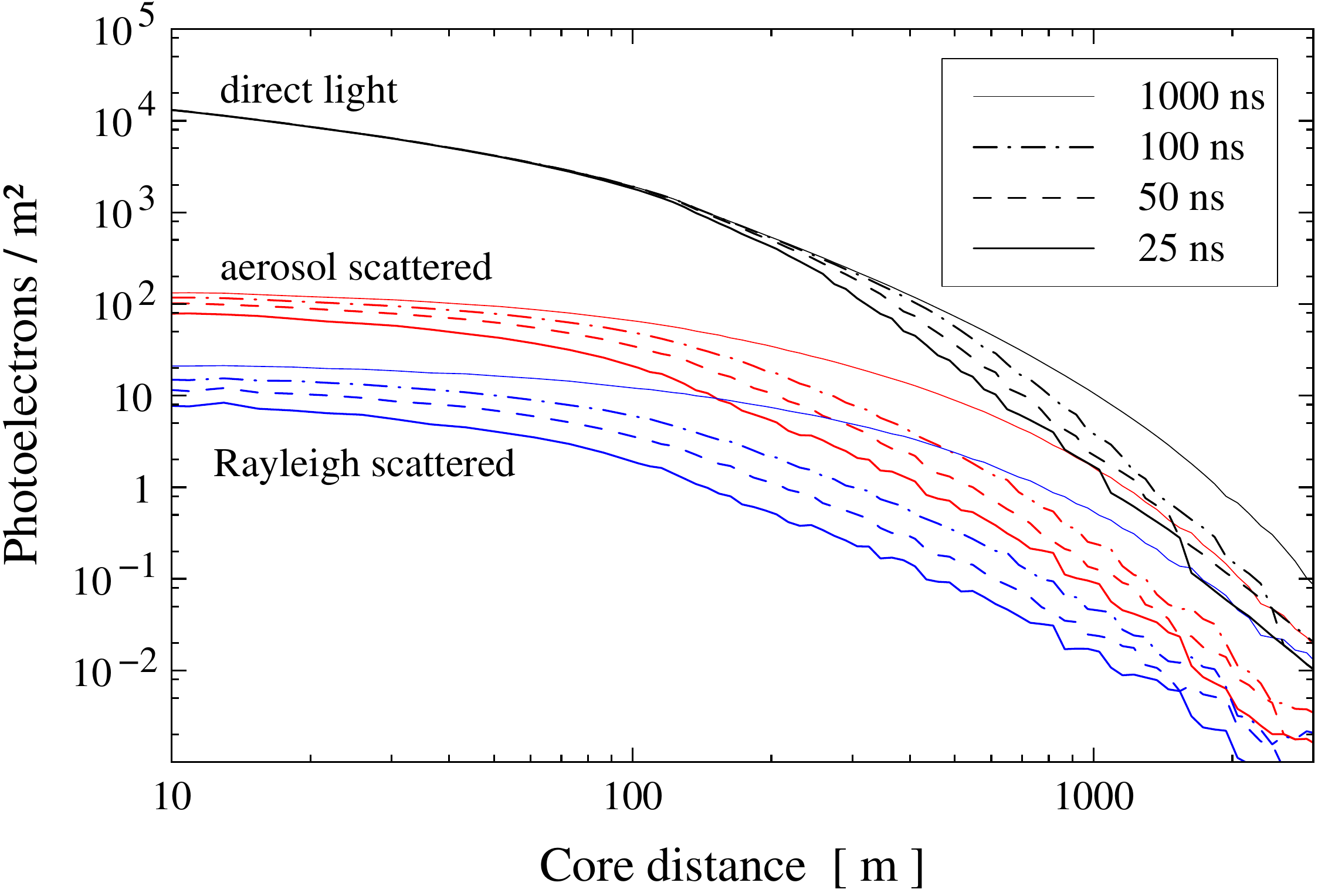}
  \caption{The relevance of scattered Cherenkov light with respect to the
  non-scattered direct Cherenkov light for different integration times,
  for vertical proton showers of 100 TeV. Mirror reflectivity and PMT quantum
  efficiency have been applied but no FoV restriction or image cleaning. 
  The aerosol scattering phase function follows a Henyey-Greenstein
  function with asymmetry parameter g=0.7 \cite{bernlohr2000}.}
  \label{fig:scattered}
\end{figure}

Both direct and scattered Cherenkov light are well known to have a
significant impact on the fluorescence technique, due to the much higher
intensity of the Cherenkov light in the forward direction, in comparison
to the weak and isotropic fluorescence light.
Scattered Cherenkov light has little impact though on the IACT technique,
due to observations being restricted to within a few hundred meters from the
shower axis, and also due to the short integration times \cite{bernlohr2000}.
Most of the scattered Cherenkov light within the camera FoV of an IACT
would arrive well outside the few nanoseconds integration time of the
signal -- only light scattered under very small angles and/or close to the
detector would have any chance of arriving within some 10 ns of the
direct light. For Rayleigh-scattered light this is of the order of $10^{-3}$
of the direct signal, for Mie-scattered light perhaps as much as a percent
(see Figure \ref{fig:scattered}).
In addition, most of the scattered light arriving within the integration time would generally
be rejected by the image cleaning used to suppress NSB noise.

Only in situations where the aerosol scattering phase function is
dominated by very large particles, with a size of about the wavelength or 
more and resulting in enhanced forward scattering, might the scattered
Cherenkov light contribution exceed the level of a few percent of the
the direct Cherenkov light. Most likely, the aerosol optical depth in
such untypical conditions would exceed levels where the IACT data is
rejected from further analysis anyway. More relevant are changes to
the scattering phase function for the fluorescence technique where
in situations of enhanced forward scattering even scattered
fluorescence light may be significant (see \cite{giller2009,pekala2009,colombi2013}).

A tricky problem to the Cherenkov and fluorescence techniques are
aerosol layers (or thin clouds) intersecting the shower. Stratospheric
aerosols from volcano eruptions in distant locations are generally no
problem since most of the shower development is below such
aerosol layers. Aerosol layers in the lower troposphere (such as the
boundary layer), i.e.\ after the end of the shower, are also not a 
major problem if properly monitored. They will affect all light in the
same way and can be easily corrected for, even from the
trigger rate of the instrument itself \cite{hahn2013},
although they will increase the energy threshold of the instrument.
The layers intersecting the shower however will change the image
shapes (for IACTs) or longitudinal profile (for fluorescence telescopes).
This will not only affect the energy calibration and detection area
but also the gamma-hadron discrimination to the extent that the
gamma-ray efficiency is very poorly known.
Although it can in principle be corrected for, if very well monitored
and simulations include these layers, the required effort in correcting
such data will generally be considered too high -- unless affecting
observations of extreme interest -- and the affected data will most
likely be discarded. 

\section{Conclusions}

Knowledge of the atmospheric density profile is important for proper
simulation of any type of ground-based air shower instrument, for
particle detector arrays as well as for Cherenkov or fluorescence detectors.
For the latter two the profile enters also in the light emission processes.
It is particularly relevant for the Cherenkov emission where the related index
of refraction determines the minimum velocity for Cherenkov emission, the
number of photons emitted per unit path length and also the Cherenkov cone
opening angle. Most simulations assume a wavelength-independent
index of refraction but the wavelength dependence can be turned on,
at the expense of the simulation efficiency.

The extinction of Cherenkov and fluorescence light depends in addition to
Rayleigh scattering and absorption on O$_2$ also on aerosols and on
trace gases like ozone and water vapour. The tropospheric ozone absorption
is of particular concern for Cherenkov detectors with extended UV response
using muon rings for calibration. Hardly any simulations make use of
measured ozone profiles.
Except for this ozone part, the atmospheric transmission tables used in
the simulations -- in particular the total aerosol optical depth -- can be
best checked with star light extinction.  For further improving the accuracy of the
simulations, the aerosol profile along the line of sight should be monitored.
Nevertheless tricky are aerosol layers intersecting the shower development,
requiring dedicated and well adapted simulations for correcting their effect.

Scattered light, in particular scattered Cherenkov light, is highly relevant to
the fluorescence technique but only has a small effect for IACTs. For IACT
simulations, it is generally ignored.

\end{document}